# Physics Opportunities with Future Proton Accelerators at CERN

A. Blondel [a], L. Camilleri [b], A. Ceccucci [b],
J. Ellis [b], M. Lindroos [b], M. Mangano [b], G. Rolandi [b]

[a] *University of Geneva*
*CH-1211 Geneva 4, SWITZERLAND*

[b] *CERN,*
*CH-1211 Geneva 23, SWITZERLAND*

**Abstract**

*We analyze the physics opportunities that would be made possible by upgrades of CERN's proton accelerator complex. These include the new physics possible with luminosity or energy upgrades of the LHC, options for a possible future neutrino complex at CERN, and opportunities in other physics including rare kaon decays, other fixed-target experiments, nuclear physics and antiproton physics, among other possibilities. We stress the importance of inputs from initial LHC running and planned neutrino experiments, and summarize the principal detector R&D issues.*

## 1 Introduction and summary

In our previous report [1], we presented an initial survey of the physics opportunities that could be provided by possible developments and upgrades of the present CERN Proton Accelerator Complex [2,3]. These topics have subsequently been discussed by the CERN Council Strategy Group [4]. In this report, we amplify and update some physics points from our initial report and identify detector R&D priorities for the preferred experimental programme from 2010 onwards.

We consider experimentation at the high-energy frontier to be the top priority in choosing a strategy for upgrading CERN's proton accelerator complex. This experimentation includes the upgrade to optimize the useful LHC luminosity integrated over the lifetime of the accelerator, through both a consolidation of the LHC injector chain and a possible luminosity upgrade project we term the SLHC. In the longer term, we consider possible future energy increases of the LHC, projects we term the DLHC and TLHC. The absolute and relative priorities of these and high-energy linear-collider options will depend, in particular, on the results from initial LHC runs, which should become available around 2010.

We consider providing Europe with a forefront neutrino oscillation facility to be the next priority for CERN's proton accelerator complex, with the principal physics objective of observing CP or T violation in the lepton sector. The most cost-effective way to do this – either a combination of superbeam and β-beam or a neutrino factory using stored muons – is currently under study by an International Scoping Study. It will depend, in particular, on the advances to be made in neutrino oscillation studies over the next few years. In the mean time, R&D is needed on a range of different detector technologies suited for different neutrino sources.

Continuing research on topics such as kaon physics, fixed-target physics with heavy ions, muon physics, other fixed-target physics and nuclear physics offers a cost-effective supplementary physics programme that would optimize the exploitation of CERN's proton accelerators. In particular, we advocate a continuing role for CERN in flavour physics such as a new generation of kaon experiments, whose topicality would be enhanced if the LHC discovers new physics at the TeV scale. However, we consider that these topics should not define the proton accelerator upgrade scenario but rather adapt to whichever might be preferred on the basis of the first two priorities. We provide below a preliminary estimate of the resources needed for R&D and a next-generation programme of kaon experiments.

## 2 LHC upgrades

### 2.1 The physics case

*Expected initial LHC results*
A successful startup of LHC operations is expected to lead rather soon to important discoveries, which will provide the required input to plan the future developments in particle physics, and to outline the optimal upgrade path for the LHC. Approximately 5 fb$^{-1}$ of integrated luminosity should be sufficient to allow the observation of a Standard Model Higgs boson over the full mass range of 115–1000 GeV, as shown in the left panel of Figure 1 [5].

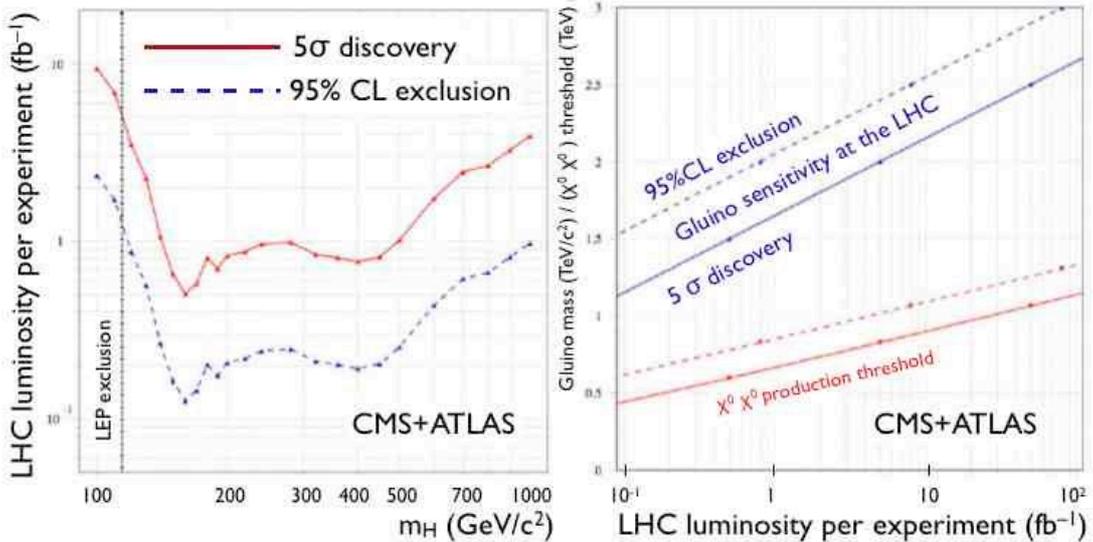

*Figure 1. Limits achievable at the 95% CL, and 5σ discovery reach, as a function of mass and integrated luminosity, for a SM Higgs boson (left panel) and for a gluino (right panel). In the latter panel, we also show the corresponding threshold for*

*sparticle pair production in electron–positron collisions, assuming universal soft supersymmetry breaking at a high unification scale [5].*

Already with 1 fb$^{-1}$ the existence of the Higgs boson anywhere within the full range of possible masses could be excluded. In parallel, 10 fb$^{-1}$ will allow the LHC to explore gluino masses up to about 2 TeV, or to set limits up to 2.5 TeV (right panel of Fig. 1). The results of these searches will set the agenda for subsequent phases of LHC exploitation, as well as determine the prospects for measuring supersymmetry directly at a 1-TeV linear collider, as also seen in the right panel of Fig. 1. A few fb$^{-1}$ would also be sufficient to detect the existence of new neutral Z′ gauge bosons up to masses of several TeV, as can be inferred from Fig. 2 [6].

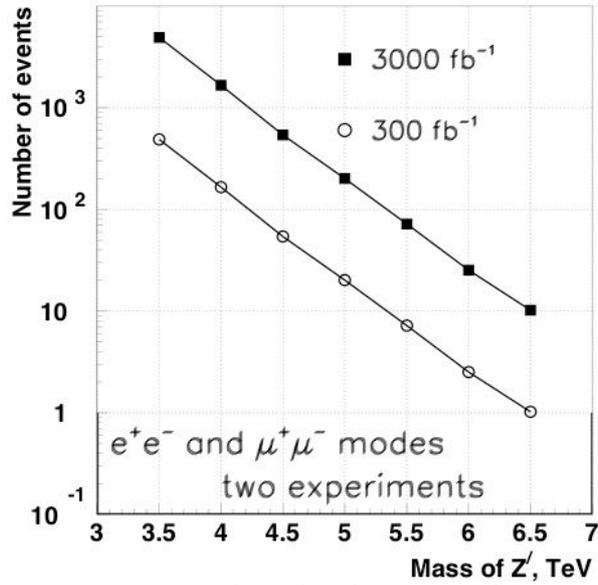

*Figure 2. Number of events expected in the electron and muon decay modes for a Z′ boson with SM-like couplings, as a function of mass and integrated luminosity. In this mass range no SM backgrounds are expected [6].*

*Full exploitation of the LHC*
The results of these and other searches for new physics in the initial runs of the LHC will strongly influence the relative priorities of LHC upgrades and linear-collider options. Their prospects should also be assessed in the light of the results that could be expected from the baseline programme of the LHC with its nominal luminosity of $10^{34}$ cm$^{-2}$s$^{-1}$, providing an integrated luminosity of several hundred fb$^{-1}$.

In the case of the Higgs boson, this includes the determination of its couplings to the fermions and to gauge bosons, as well as establishing the scalar nature of this state. The left panel of Fig. 3 displays the accuracies obtainable at the LHC in measurements of the couplings of a Higgs boson weighing 120 GeV [7,3]. They would enable a demonstration that many Higgs couplings are proportional to particle masses at the level of about 20%. Some of these determinations rely on an assumption that the Higgs has no substantial invisible decay modes. This assumption may be justified by recent preliminary studies, which have shown that the LHC experiments will also be sensitive to invisible Higgs boson decays via several production channels, namely in association with the Z or a top–antitop pair, and in vector-boson-fusion [8]. With an integrated luminosity of less than 30 fb$^{-1}$, each experiment should be able to establish a 5σ signal for a Higgs boson with a 100% branching ratio for invisible decays; with 10 fb$^{-1}$ of integrated luminosity, to establish a 95% confidence-level upper limit on the invisible branching ratio as low as 15 to 30%, for any Higgs boson mass between 115 and 400 GeV [8]. The spin of a Higgs candidate is specified if the resonance is seen to decay into gamma–gamma or ZZ, which are among the favoured decay modes for a Higgs boson in the Standard Model, as in that case it cannot have spin 1. The Z*Z decay mode also provides discrimination between scalar and pseudoscalar decay modes [9].

In the case of supersymmetry, the nominal LHC luminosity will provide access to a large range of possible sparticle masses [10], as also seen in the right panel of Fig. 1, providing a 5σ discovery potential for a gluino weighing up to 2.7 TeV and 95% CL exclusion up to 3.2 TeV. In the event that sparticles are relatively light and have been detected already in initial LHC runs, the additional luminosity will enable many sparticle species to be observed, and permit interesting measurements of their masses and couplings [11]. These measurements might also make possible a first calculation of the astrophysical dark-matter density using collider data.

In the case of new gauge interactions, their couplings to SM particles will need to be determined. As indicated in Figs. 1 and 2 above, increasing luminosity by a factor of 10 will also lead to increasing the sensitivities to new particles at the level of 20–30% in the mass reach. The amount of additional statistics required to saturate the systematic uncertainties in the precision of the measurements will depend on the detailed features of what is observed, in particular on the mass and production rates.

*LHC upgrade scenarios*
Upon completion of this baseline programme, two possible scenarios for LHC upgrades are conceivable: a further increase in luminosity, and an increase in energy. In the first case, what is discussed is an increase of the luminosity to $10^{35}$ cm$^{-2}$s$^{-1}$ (a project known as the SLHC). In the second case, the possibilities of doubling (DLHC) [12] or even tripling (TLHC) [13] the nominal beam energy of 7 TeV have been proposed. We discuss first the physics cases for these options, and then review the implications for the accelerator and for the experiments.

*SLHC physics*
Higher LHC luminosity will benefit a large number of Standard Model measurements that we already know will be possible, for example the study of rare top decays and the determination of the self-interactions of gauge bosons. Many examples of the additional new physics accessible via a tenfold increase in the LHC luminosity, to $10^{35}$ cm$^{-2}$s$^{-1}$ (SLHC), were given in [6]. They include:

- **Higgs physics** – Improved determination of the Higgs-boson couplings to fermions and gauge bosons; observation of rare Higgs decays as H to Z$\gamma$ and H to $\mu\mu$, detection of Higgs pair production and measurement of the Higgs self-coupling (see the right panel of Fig. 3).
- **Strong gauge-boson scattering** – In the case that no Higgs boson is observed in the baseline LHC programme, the study of strongly coupled vector bosons will be required, and hence large luminosity. The example of a 750 GeV resonance in the $Z_L Z_L$ channel is shown in Fig. 4. As is clear from the rates, this measurement would not be possible at nominal luminosity.
- **Electroweak measurements** – Improved measurements of multiple-gauge-boson production will be possible, enabling precise measurements of the triple-gauge-boson couplings to be made at the level of electroweak radiative corrections (see Fig. 5).
- **Supersymmetric Higgs sector** – The discovery of the additional Higgs bosons predicted by supersymmetry is challenging for the nominal LHC programme, particularly for large masses of the CP-odd Higgs $A^0$. The SLHC can increase by over 100 GeV the discovery range in the most difficult region of $\tan\beta \sim 5$, fully covering the domain of sensitivity of measurements at the ILC (see Fig. 6).
- **Searches for new physics** – The SLHC would enable the mass reach for discovering gluinos to be extended from about 2.7 TeV to about 3.3 TeV (see Fig. 7), covering a much larger fraction of the parameter space where supersymmetry could provide astrophysical dark matter. The SLHC would also enable the mass reach for a new Z′ to be extended from about 5 TeV to about 6 TeV, and the scale for compositeness from about 30 to 40 TeV.

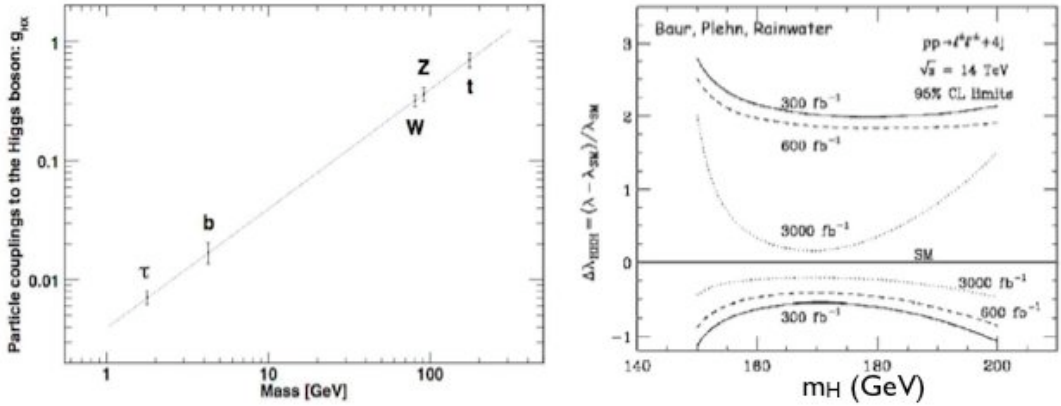

*Figure 3. Determination of the Higgs couplings (left panel) and (right) limits achievable at the 95% CL on the deviation of the triple-Higgs coupling from the Standard Model value, $\lambda_{HHH}$, at the LHC and SLHC [14]. The allowed region is between the two lines of equal texture. The Higgs boson self-coupling vanishes for $\lambda_{HHH} = -1$.*

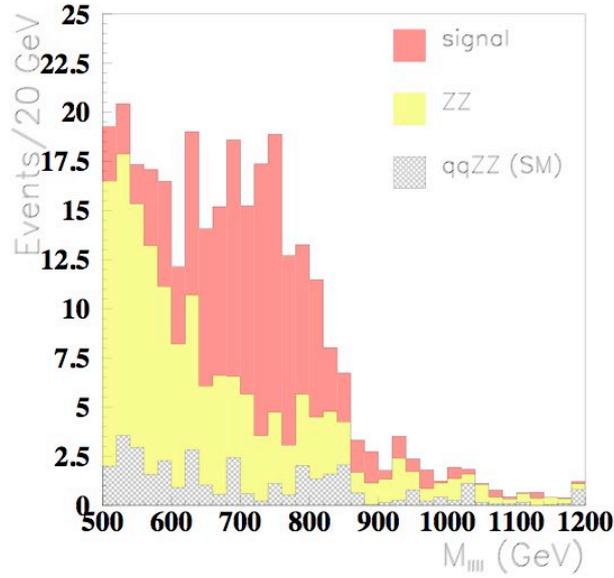

*Figure 4. Expected signal and background at the SLHC (3000 fb$^{-1}$) for a scalar resonance of mass 750 GeV decaying into four leptons [6].*

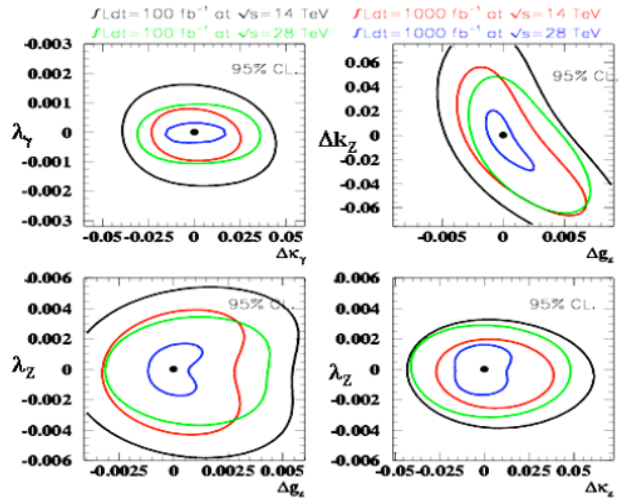

*Figure 5. Expected 95% CL constraints on triple-gauge-boson couplings [6]. The black contours correspond to 14 TeV and 100 fb$^{-1}$ (LHC), the red to 14 TeV and 1000 fb$^{-1}$ (SLHC), the green to 28 TeV and 100 fb$^{-1}$, and the blue to 28 TeV and 1000 fb$^{-1}$ (DLHC).*

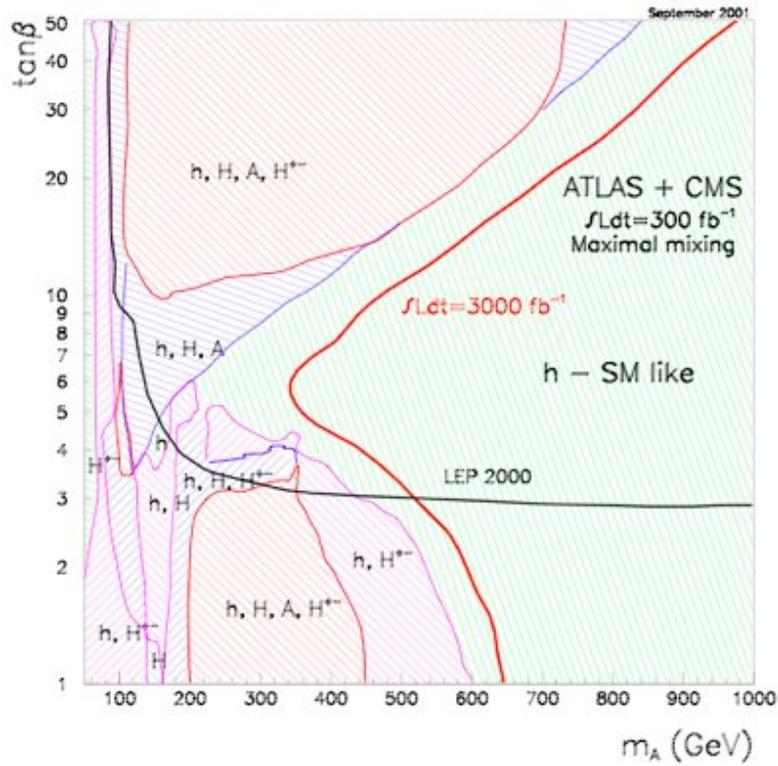

*Figure 6. Regions where the various MSSM Higgs bosons can be discovered at the LHC. The SLHC extends the domain for discovery of the heavier MSSM Higgs bosons to the rightmost red line [6].*

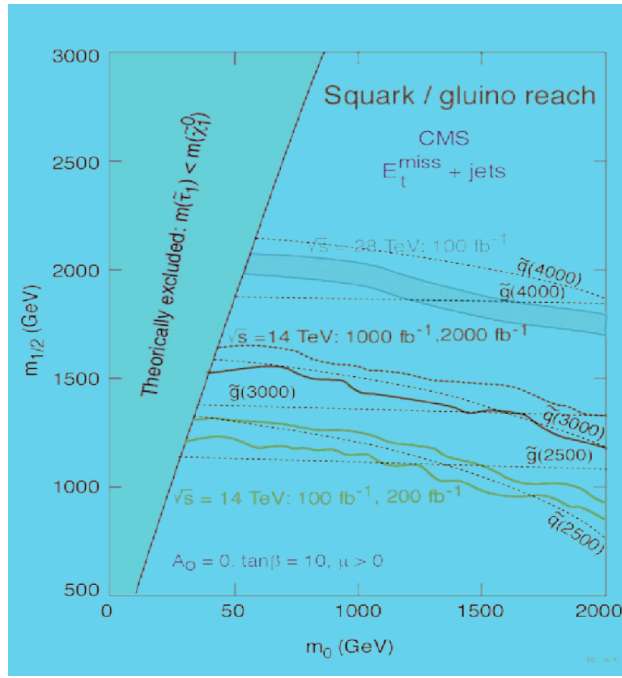

*Figure 7. Expected 5σ discovery contours for supersymmetric particle masses in a typical model plane [6]. The various curves show the potentials of the LHC for luminosities of 100 fb$^{-1}$ and 200 fb$^{-1}$, of the SLHC for 1000 fb$^{-1}$ and 2000 fb$^{-1}$, and of the DLHC for 100 fb$^{-1}$.*

*DLHC and TLHC physics*

An increase in energy is a natural upgrade path to pursue further the search for very massive new particles and to explore the shortest possible distance scales. Aside from the generic desire for higher energy, several direct or indirect observations at the LHC could make this need more explicit. For example:

- LHC discovers new particles at the edge of the kinematically allowed phase space (e.g. gluinos above 2.5 TeV, or a Z′ above 5 TeV). In this case an increase in energy would enhance the production rates, and the potential for a quantitative study of the new particles, much more than a luminosity upgrade.
- LHC observes a departure from the point-like behaviour of quarks, via an anomaly in the high-$E_T$ jet spectrum. Only a substantial increase in energy will allow a direct study of the quark substructure.
- Phenomena like extra dimensions, or a strongly-interacting Higgs sector, are revealed by the LHC. In both cases, higher energy is required. For example, in the case of extra dimensions, the observation at the LHC of the first Kaluza–Klein excitations of ordinary particles would require access to higher energies, to verify the expected tower structure of massive states, and to get information on the nature of the extra dimensions. In the case of a strongly interacting Higgs sector, such as the completion of little Higgs theories, the low-lying modes of the theory, expected to lie below 5 TeV, could be accessible at the LHC, but the scale of the new strong interactions is expected to lie in the range of 5–10 TeV. This is well beyond the LHC reach, regardless of its luminosity, and the DLHC or TLHC would be the only way to explore these scenarios further.

The above points can be illustrated with some specific examples. The following figure shows the production cross section for pairs of heavy quarks at the three different centre-of-mass energies of 14, 28 (DLHC) and 42 TeV (TLHC), as a function of the quark mass. This example applies, in particular, to searches for the singlet T-quark partner of the top quark in little-Higgs models [15].

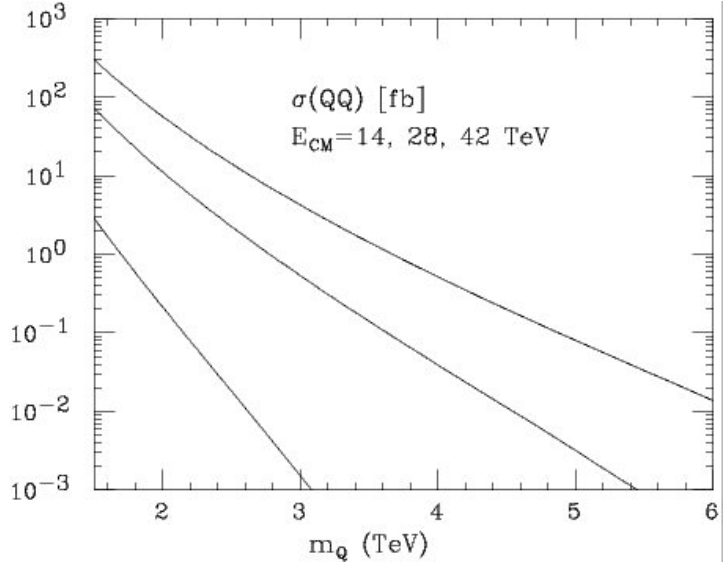

*Figure 8. Cross sections (in fb) for production of heavy quark pairs. Lines of increasing cross-section correspond to pp collisions at centre-of-mass energies of 14, 28 (DLHC) and 42 TeV (TLHC).*

We notice that already for quarks weighing 1.5 TeV, whose cross section at the LHC is in the range of a fb, a doubling in energy is statistically more effective than a tenfold increase in luminosity. In the case of gluino production, the production rates are about one order of magnitude larger, but the relative rates at the different energies are similar to the case of heavy quarks. The DLHC would essentially complete the coverage of the parameter space for supersymmetric dark matter.

In the case of new gauge interactions, Fig. 9 shows, as an example, that for a W' with mass above 3.5 TeV the doubling of energy is a more effective way of increasing the statistics than an order-of-magnitude increase in luminosity. We also include the comparison with the TLHC, which would allow one to push the W' mass reach beyond the 10 TeV range at a luminosity of $10^{34}$ fb$^{-1}$.

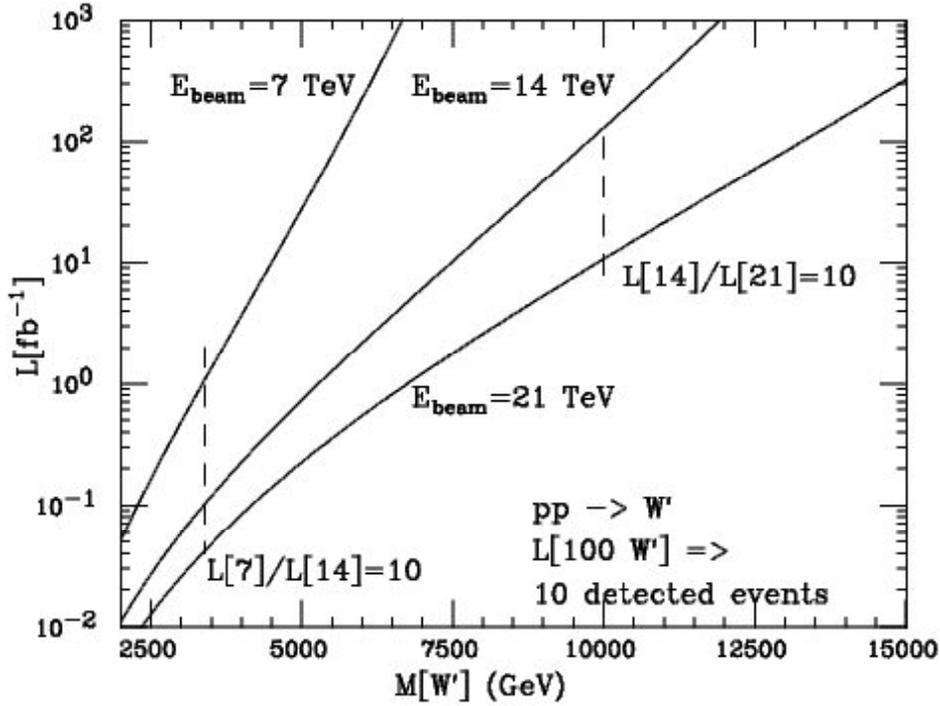

*Figure 9. A sample comparison between the SLHC, the DLHC and the TLHC: at low mass ~ 2 TeV, the W' cross section increases by a factor of 5 at the DLHC and another factor at the TLHC, but the increase is much larger for a heavier W'. The DLHC would require only a few $fb^{-1}$ to discover a W' weighing 7 TeV, which would require 1000 $fb^{-1}$ at the SLHC. The reach at the TLHC would extend beyond 10 TeV.*

Table 1 shows a brief summary of the comparison of the physics potentials of the basic LHC and the SLHC. We also give some examples of the physics reach of the DLHC and, for comparison, the physics reaches of an electron–positron linear collider of 0.8 TeV and CLIC with 5 TeV.

*Table 1: The reach of LHC, SLC and DLHC are compared with those of a linear collider of 0.8 TeV and CLIC at 5 TeV [6].*

| Process | LHC 14 TeV 100 $fb^{-1}$ | SLHC 14 TeV 1000 $fb^{-1}$ | DLHC 28 TeV 100 $fb^{-1}$ | LC 0.8 TeV 500 $fb^{-1}$ | CLIC 5 TeV 1000 $fb^{-1}$ |
|---|---|---|---|---|---|
| Squarks (TeV) | 2.5 | 3 | 4 | 0.4 | 2.5 |
| $W_L W_L$ ($\sigma$) | 2 | 4 | 4.5 | 6 | 90 |
| Z' (TeV) | 5 | 6 | 8 | $8^\perp$ | $30^\perp$ |
| Extra-dimens. scale (TeV) | 9 | 12 | 15 | $5–8.5^\perp$ | $30–55^\perp$ |
| q* (TeV) | 6.5 | 7.5 | 9.5 | 0.8 | 5 |
| Compositeness scale (TeV) | 30 | 40 | 40 | 100 | 400 |
| TGC, $\lambda_\gamma$ (95%CL) | 0.0014 | 0.0006 | 0.0008 | 0.0004 | 0.00008 |

$^\perp$*Indirect reach from precision measurements*

## 2.2 Accelerator challenges

Achieving the nominal LHC luminosity of $10^{34}\,\text{cm}^{-2}\text{s}^{-1}$ with high reliability and efficient operation is first priority for the physics programme and for the correct exploitation of the large investment on LHC. However, this may well prove challenging, particularly in view of the ageing of the PS and the SPS. For this reason, we support efforts and investment of resources to consolidate the LHC injector complex [2,3].

We expect that the LHC luminosity will increase gradually with time, thanks to experience in its operation, incremental hardware improvements and consolidation. As shown above, initial lower-luminosity running will already provide useful inputs for possible upgrades beyond the nominal LHC performance. The luminosity may eventually reach a factor of 2 above the nominal luminosity, if the beams collide only in IP1 and IP5 and the bunch population is increased to the beam–beam limit.

Increasing the LHC luminosity above this figure would require hardware changes in the LHC insertions and/or in the injector complex [2,3], as well as necessitate substantial modifications to the LHC detectors. We note in passing that an 8% increase in the LHC energy, which might be possible with the available LHC magnets, would permit improved studies of new heavy particles. We recognize that doubling or even tripling the LHC energy would be much more challenging projects, with longer R&D lead times and substantially higher costs.

## 2.3 Detector upgrades

Experimentation at the SLHC will be more difficult than at the design luminosity of the LHC, because of the large increase in pile-up – by a factor of 5 to 10, depending on the interval between bunch crossings – and of the large irradiation of the detectors. The physics reach will be the result of an optimization between the increase in integrated luminosity and the more challenging running conditions that will strain the performance of the detectors [6]. The first LHC runs will give us input on some parameters that are needed for the design of the upgraded SLHC detectors, such as neutron fluence, ageing, radiation damage, activation and performances of the present detectors.

The present inner tracking systems are designed to survive a maximum of about 300 $\text{fb}^{-1}$, after which they will in any case need to be replaced by new devices, during a long shut-down. In the event of a luminosity upgrade, they should also be capable of sustaining the increase in pile-up, and hence will need more detector elements to deal with the occupancy levels. Electronic technology evolution will bring benefits and should be adopted, but the associated power distribution is an issue requiring further study, as well as the integration of services such as cooling in the existing space. Radiation-hard silicon sensors are being developed in the framework of the RD50 collaboration, in conjunction with industrial partners.

One important ingredient for the luminosity increase is the modification of the insertion quadrupoles to yield a $\beta^*$ of 0.25 m, compared with the nominal 0.5 m. The new interaction regions have yet to be defined, and their layout may have significant implications for the experiments. Studies of the various options should be pursued aggressively, using as a basis concrete examples of a mechanical layout and envelopes of the elements as well as of the services necessary for their operation.

Another important ingredient for the luminosity increase and for the reduction of the pile-up is the reduction of the spacing between beam crossings at the SLHC. At present it is not clear if this reduction is feasible, in view of the electron-cloud effect and the thermal load on the cryogenic system. However, the upgrade of the electronics of the calorimeters and of the muon systems would depend strongly on this new spacing, and the LHC experiments have expressed clear preferences for going to a spacing of 12.5 ns (one half of the present 25 ns), which could allow most of the front-end electronics for the calorimeter and muon system to continue running at 40 MHz. A spacing of 10 or 15 ns – which would avoid changes to the timing of the SPS – would be likely to require much more complex modifications to the front-end electronics of these subsystems. In the case of the tracking detectors, new front-end electronics would be designed according to the new selected bunch spacing. The planning of the R&D and of the upgrade of the front-end electronics depends crucially on the bunch-crossing frequency; it is important that issues related to the reduction of the bunch spacing are clarified experimentally during the first LHC runs, if possible before the end of 2008.

The technical challenges are already under study by the ATLAS and CMS Collaborations, together with the modifications to the Trigger and DAQ systems and to the radiation protection and shielding that would be required in any SLHC scenario. Among the common R&D topics identified by the collaborations are the inner-detector readouts, ASICs, radiation-hard silicon sensors and modules, radiation and activation, bump bonding, power management and cooling, optoelectronics and control links. The R&D needed for the upgrades of the detectors has synergies with the EUDET program and ILC detector R&D in general, especially for items linked to the likely evolution in electronic technology.

The largest fractions of the upgrade cost are due to the replacement of the trackers, which would in any case be needed after about 300 $fb^{-1}$. The R&D cost is evaluated to be around 10% of the capital cost. The industrial production and qualification of many hundreds of square meters of silicon is a challenge that should be addressed in a timely manner. In most cases, the replacement of the on-detector electronics will require a complex and manpower-intensive procedure for accessing the electronics itself.

Both collaborations have already set up management structures to coordinate and prioritize the required R&D projects. Several groups around the world are getting involved in R&D programmes on upgrades related to their expertise and instrumentation interests, and they are generally ahead of CERN. This is quite natural, as many outside groups now have personnel available at their institutes, following the construction of their deliverables, while the installation and commissioning effort at CERN will go on strongly and with the highest priority for another 2 years or so.

As emphasized earlier, the definition and schedule of the preferred LHC luminosity upgrade scenario in 2010 will depend on results from the initial LHC runs, including physics results and assessment of the LHC running conditions. Meanwhile, we encourage any measure to achieve and improve the performance of the LHC in a timely manner, including the ultimate beam parameters and the highest possible energy.

## 2.4 Final remarks on the energy upgrade

We recognize that either the DLHC or TLHC accelerator would require a substantial long-term investment in developing new high-performance magnets, in order to increase the bending field. There should be synergy with the R&D on high-field magnets for the SLHC collision insertions. It is worth noticing that, whereas the upgrade from 14 TeV to either 28 or 40 TeV would represent a major leap in discovery reach, the step from 28 to 40 TeV, being only a 50% increase, is not as effective in relative terms. This is shown in some of the examples provided above, in the discussion of the physics case. The implication of this observation is that, given the high costs and the scale of these upgrades, it is unlikely that a TLHC would follow the DLHC. We should therefore ideally aim at establishing the feasibility of both projects within a comparable time-scale, to decide which of the two jumps to focus on.

We note that some of the scenarios for upgrading CERN's proton accelerator complex would have particular relevance if either the DLHC or the TLHC were to be envisaged: for example, a higher-energy replacement for the SPS would be required [2,3].

The detector R&D needed for the DLHC is smaller than that required for the SLHC, and one could in principle use ATLAS and CMS with only small modifications. However, more R&D would be needed for experiments at the TLHC, in particular on topics such as the momentum up to which charge separation is possible, and the impact of calorimeter leakage on the missing transverse energy. Moreover, we recognize that on the time scale of a DLHC or TLHC project the present detectors will require the replacement of aged/old subdetectors using the new technologies, e.g. electronics and DAQ, that will be available then. There should be synergy with the detector R&D for the SLHC.

## 2.5 ALICE upgrades [16]

The ALICE experiment at the LHC envisages a comprehensive programme of physics that is expected to last until around 2017, and hence should be taken into account in planning for the SLHC. We note in passing that the approved RHIC programme is expected to terminate around 2012, after which ALICE will be the only high-energy heavy-ion collider experiment. ALICE plans to explore different beam types and energies as well as increase the luminosity as far as possible. The first Pb–Pb run is expected in 2008, to be followed by further Pb–Pb runs in the following two running years, at least. ALICE also plans continuous proton–proton running and requests some proton–ion running in 2010 or later, which would be necessary to isolate the collective effects in ion–ion collisions better. Running with low-mass ions and at lower energies are also requested for later years.

The collaboration plans to complete its baseline detector by 2010, including the PHOS, TRD and time-of-flight systems. These may require some access to test beams and manpower for electronics integration. It is also planned to install by 2010 an electromagnetic calorimeter, with no significant cost to CERN. Further detector upgrades are foreseen for construction in 2010–2013 and installation in 2012–2015. These would include a thinner beam pipe closer to the interaction point and new pixel detectors. There are also plans for improved, higher-transverse-momentum particle identification and improved instrumentation for forward physics.

ALICE also requests an accelerator R&D programme to increase the Pb–Pb luminosity by a significant factor, aiming at $5\times10^{27}$ cm$^{-2}$s$^{-1}$. This would be required to make good measurements of Y, Y' and Y'' production as well as γ–jet correlations, probes at the highest possible transverse momentum, and beauty production. In order to gather sufficient statistics, the ALICE Collaboration would like to run for three or four years at this enhanced luminosity. Obtaining this would require fighting pair production and electron capture, and rapid turnarounds and refills would also be desirable. In view of the lower multiplicity now expected in the TPC, its readout could be increased to cope with the higher event rate. Modifications to the TPC itself, to its electronics and to the DAQ system would be needed.

## 2.6 LHCb upgrades [17]

The following are some examples that demonstrate the physics reach and topicality of LHCb. With its nominal integrated luminosity of 10 fb$^{-1}$, LHCb would be able to measure the angle γ with a precision of below 5 degrees, make a first measurement of the $B_s$ oscillation phases, and be sensitive to $B_s \to \mu\mu$ decay at the level of the Standard Model expectations. Also, 22000 $B_d$ to $K^{*0} \mu^+\mu^-$ events would be collected, a number far greater than what can be achieved by the B factories and which would allow an accurate measurement of forward–backward asymmetries that might be sensitive to new physics.

The LHCb programme is not limited by the LHC luminosity. Initially, LHCb will run at a luminosity of $2 \times 10^{32}$ cm$^{-2}$s$^{-1}$, which is limited by the desire to minimize multiple interactions and by the radiation dose in the VELO silicon tracker, which is located within 8 mm of the beam. Given the current estimate of the running time available per year, the integrated luminosity required by LHCb to attain its physics goals, namely 10 fb$^{-1}$, would require a 10-year programme at an average luminosity of $2 \times 10^{32}$ cm$^{-2}$s$^{-1}$.

The required running time could, however, be reduced by operating the experiment at higher luminosity. After understanding the detector, it seems possible that the luminosity could be increased gradually to a level of $5 \times 10^{32}$ cm$^{-2}$s$^{-1}$. This should increase the number of events obtained via the muon trigger by a factor of 2, but with no corresponding increase in the number of hadronically triggered events, because of the less selective nature of the hadronic trigger. During this time the detector upgrades would be modest, if any. One may replace the VELO sensors so that they can be moved to within 5 mm of the beam by improving the guard-ring design, or study the possibility of removing the RF foil, which would improve the impact parameter resolution by 36%.

As a second step to speed up the programme significantly, and potentially increase its scope, LHCb would attempt to run at a luminosity of $10^{33}$ cm$^{-2}$s$^{-1}$, assuming that multiple interactions per crossing can be managed. The physics motivation of this upgrade would have to be confirmed by the physics landscape in coming years. If ATLAS and CMS discover new physics at the TeV scale, LHCb could measure its couplings indirectly in different channels, helping in the identification of the new physics and the exploration of its flavour structure. Alternatively, if no new signals were observed, LHCb could probe the multi–TeV energy scale indirectly through precision measurements in the flavour sector.

For such a luminosity upgrade to be useful, the first–level trigger would have to be modified so as to include an impact parameter requirement in addition to a $p_T$ requirement. For this to be possible, the VELO and TT station would have to be read out in 25 ns. New first-level electronics would have to be developed, so as to keep the latency at a level of 1.7 ms and thereby avoid replacing the whole front-end electronics. An alternative is to make use of the latest high-speed Ethernet and network technology to read all the data out at 40 MHz and transmit them to a farm. However, this would be substantially more expensive and require more time for the development of all the front-end electronics. More radiation-hard silicon sensors would be needed, such as the pixels being developed at Syracuse and FNAL. The inner modules of the electromagnetic calorimeter would also need to be replaced, in order to cope with high occupancy and irradiation while maintaining the $\gamma$, $\pi^0$ and electron identification. The outer and inner trackers, some of the muon chambers, and a small number of HPDs in RICH-1 would also need replacing, so as to increase their granularity and rate capability. Finally, the number of DAQ readout boards and the speed of the readout network would also need to be increased at this higher event rate, and the power of the CPU farm would have to be increased for processing more complex events with multiple pp interactions in one event.

A realistic R&D and construction time scale would lead to data taking in the new configuration in 2014, possibly around the time of the SLHC. The flexibility allowed by the LHCb IR magnets is such that a luminosity of $10^{33}$ cm$^{-2}$s$^{-1}$ could be achieved at LHCb even at a nominal LHC luminosity of $10^{35}$ cm$^{-2}$s$^{-1}$. However, a halving of the bunch spacing would have a severe impact on such an LHCb upgrade programme.

# 3 Neutrino physics

## 3.1 Overview

Neutrinos have provided the first clear experimental evidence for physics beyond the Standard Model, and there are many important outstanding issues to be addressed by future accelerator neutrino experiments. These include the magnitude and hierarchy of neutrino masses, the magnitude of the third neutrino mixing angle $\theta_{13}$ and the possibility of CP or T violation in neutrino oscillations, as well as the overall verification of the assumed theoretical framework. The study of neutrino mixing angles, mass differences and CP violation has a possible connection to grand unified theories through the see-saw mechanism, while leptonic CP violation provides one of the most elegant mechanisms for understanding the baryon–antibaryon asymmetry of the Universe. Neutrino physics would be even richer if the LSND experiment were confirmed.

The magnitude and hierarchy of neutrino masses will be addressed by neutrinoless double-β decay experiments, cosmological measurements and accelerator experiments on neutrino oscillations, though these issues are unlikely to be resolved before 2015. The observation and study of CP violation requires powerful and precise neutrino oscillation experiments in suitable channels. Theory currently does not provide useful guidance on the magnitude of $\theta_{13}$, which is being addressed by a series of oscillation experiments. These include OPERA and perhaps Double-Chooz in Europe, MINOS and perhaps NOνA in the US, T2K in Japan and perhaps another reactor experiment in the US, Japan, China or Brazil. As seen in Fig. 10, the world sensitivity may reach $\sin^2 2\theta_{13} \sim 10^{-2}$ by 2010–2012 after the first results from Double–Chooz and T2K, and a few $10^{-3}$ by 2015. None of the projects discussed for the next ten years has any significant sensitivity to the CP–violating phase δ in the simplest 3-generation mixing scheme.

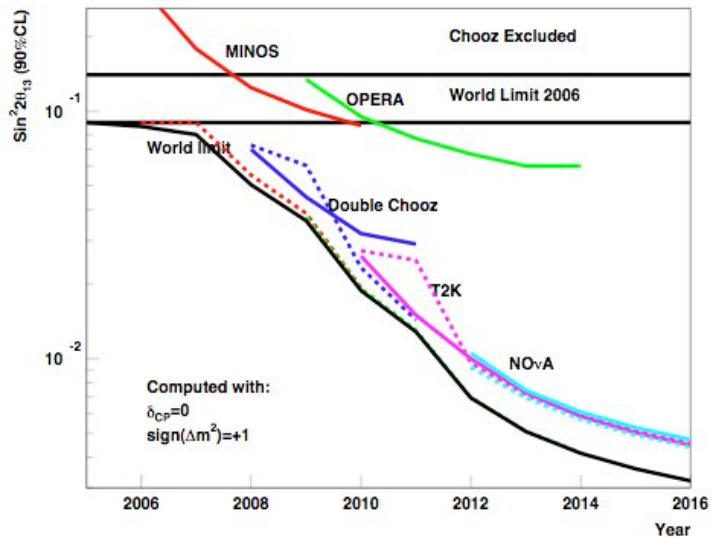

*Figure 10 – Estimate of the possible future evolution of the experimental sensitivity to $\theta_{13}$ [18].*

A programme of precision neutrino oscillation physics, leading to such a major discovery as the observation of leptonic CP violation, would be an appropriate, exciting, high-level goal for CERN. Such a high-intensity programme would provide abundant additional physics possibilities, including studies of neutrino matter effects and the mass hierarchy, precision measurements of mixing angles and precision neutrino-interaction physics. It would also offer interesting prospects for muon physics and nuclear physics, as described later in this report.

The value of the CP asymmetry depends on the value of the mixing angle $\theta_{13}$, and is maximal for a value of $\sin^2 2\theta_{13}$ of the order of $10^{-2}$ to $10^{-3}$. The CP sensitivity of proposed experiments is quite sensitive to the level of background and on systematic uncertainties. These issues are highlighted in Fig. 11, where the sensitivity reaches of three well-studied possible facilities have been compared:

i) The T2HK concept, an extrapolation of the T2K experiment with a megaton water Cherenkov detector exposed to the same ~ 600 MeV muon–neutrino beam produced by 2.5° off-axis pion decay as used for T2K, but produced with a hypothetical proton accelerator upgraded to deliver 4 MW power;

ii) A superbeam/β-beam megaton facility in which a megaton detector situated in a possible extension of the Fréjus underground laboratory is exposed to both an on-axis superbeam generated at CERN with a high-intensity proton driver of energy 3 to 5 GeV, and to a β-beam using $^6$He and $^{18}$Ne ions accelerated in the existing CERN PS-SPS complex, thus allowing exposure to beams of $\nu_\mu, \overline{\nu}_\mu, \nu_e, \overline{\nu}_e$ with energies of about 200 to 500 MeV.

iii) A neutrino factory based on a muon decay storage ring, as studied for a 4 MW proton driver of energy 5 to 30 GeV and able to deliver $6 \times 10^{20}$ useful muon decays to a 100 kiloton magnetized iron detector situated 3000 km away and a coarser-grain 30 kiloton detector situated 7000 km away.

One of the goals of the International Scoping Study launched, in particular, by the EU-funded network BENE [19] of CARE, is the evaluation of the relative merits of a neutrino factory, a superbeam and a low-γ β-beam for large values of $\sin^2 2\theta_{13}$, taking into account systematic uncertainties. Two important sources of systematic errors have been identified: for the low-energy experiments, the understanding of systematic uncertainties related to nuclear effects, Fermi motion, nuclear binding energy and nuclear pion absorption; for the high-energy neutrino factory, the understanding of matter effects plays a dominant role in the systematic errors. Recent studies show that the error associated to nuclear effects of CP asymmetries in low-energy experiments could be of the order of 6% at 250 MeV, decreasing rapidly at higher energies. However, this error needs to be validated with a careful design of the near detector station. The recent studies also indicate that the matter density needed for a neutrino factory could be known to about 2% precision.

One may draw the following conclusions from Fig. 11. If $\sin^2 2\theta_{13}$ is very near the present limit, an extrapolation of T2K may be able to see CP violation, although with remaining ambiguities. However, we note that there is no obvious path for upgrading J-PARC much beyond 1.25 MW beam power. For $\sin^2 2\theta_{13}$ larger than about $10^{-2}$, it may be possible to measure the CP–violating phase δ using the superbeam+β-beam+megaton combination, for which the increased sensitivity obtained by combining the superbeam and the β-beam is clearly essential. On the other hand, the neutrino factory with distant magnetized detectors at one or possibly two very long baselines would be able to measure δ and determine unambiguously the oscillation parameters for a much larger domain of parameter space. Having a higher energy allows, in particular, great sensitivity to matter effects, and, if a suitable detector can be built, unique observation and study of $\nu_e$ to $\nu_\tau$ oscillations would be possible.

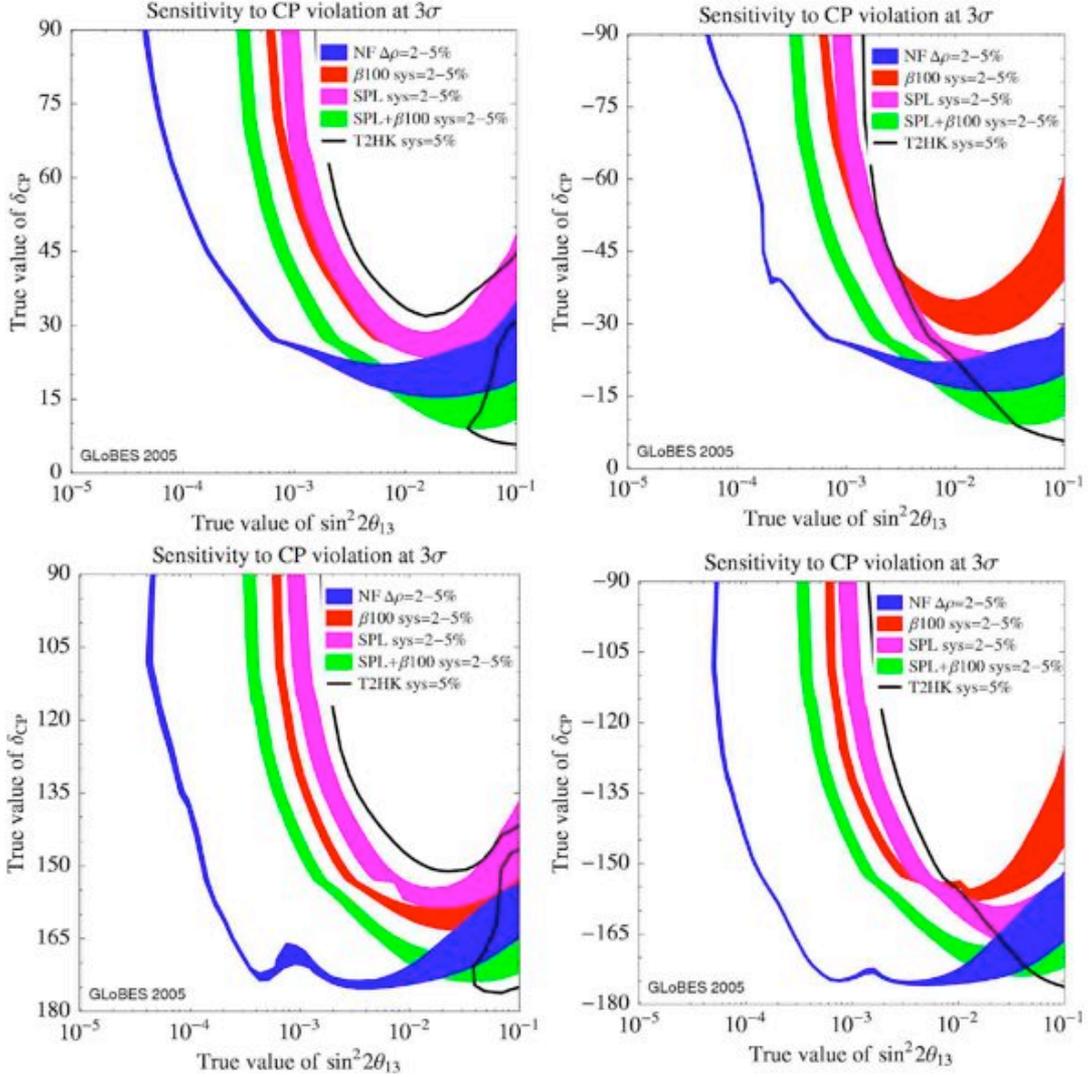

*Figure 11 – Provisional estimates of the sensitivities of various proposed neutrino facilities to the CP-violating phase δ in the simplest three-generation model of neutrino mixing [20]. The black line assumes that J-PARC delivers 4 MW for 8 years to Hyper-Kamiokande, a megaton water Cerenkov detector. The magenta line is the SPL superbeam running for 10 years aimed at a 440 kiloton water Cerenkov detector at a distance of 130 km, the red line is the γ = 100 β-beam aimed at the same detector, and the green line is a combination of the two. The blue line is the neutrino factory optimized for small values of $\theta_{13}$, aiming at detectors 3000 and 7000 km away. The thickness of the lines shows the effect of varying the systematic errors related to cross sections or matter effects within the indicated ranges. A definitive version of this comparison is a deliverable of the International Scoping Study [21].*

The competition in the search for leptonic CP violation after 2015 is not yet clear, but projects are being discussed actively in the US, including programmes based on upgrades of the proton accelerator complexes at FNAL or BNL, and, in Japan, namely a successor to T2K based on an upgrade of JPARC to a few MW in conjunction with one or two megaton water Cerenkov detectors, one in Japan (Hyper-Kamiokande) and one perhaps in Korea. The prospects of these projects will be determined by the results obtained in the first round of experiments, as well as by the time scale and location of the ILC, but it seems unlikely that Hyper-Kamiokande could be completed before about 2025. It therefore appears that a high-power (4 MW) proton accelerator at CERN would open the possibility of an extremely competitive neutrino programme during the next decade.

It should also be re-emphasized that many more exciting vistas in neutrino physics would open up if the results of the LSND experiment are supported by the ongoing MiniBooNE experiment, which will soon announce its first results.

These conclusions must be taken with great care. First, it is clear that all the suggested facilities pose considerable challenges, and that practicality and feasibility need to be determined by a careful design study. Secondly, not all possibilities have been explored and new ideas are to be expected.

For instance, a higher-energy β-beam option (He/Ne with $\gamma = 350$, or using high-Q decay isotopes) with longer baselines would be sensitive to values of $\sin^2 2\theta_{13}$ smaller than the conventional one for He/Ne $\gamma=100–150$, but would require an upgrade of the SPS to higher energy, or acceleration of the radioactive ions in the decay ring. In the latter case, it has been suggested that this post-acceleration capability could also be used to increase the energy of particle injection into the LHC.

Another interesting idea is the generation of a monochromatic neutrino beam in a β-beam facility using nuclei decaying via electron capture. The main challenge is the long half-lives of most nuclei for which electron capture is responsible for a large fraction of the decays. The recent discovery of nuclei in the rare-earth region (e.g. $^{152}$Tm), with half-lives as short as 8 sec decaying mainly via electron capture, due to a Gamow–Teller resonance in the daughter nucleus, has triggered a study of the physics reach for a monochromatic β-beam. A first study of the machine parameters for such a β-beam indicates that the half-life is still too long and the projected production rate too low to generate a sufficient yield of neutrinos. The annual rates estimated at the moment are too low for a CP-violation measurement, although they could be useful for a near-detector calibration and cross-section experiment. However, possible advances in target and ion-source systems (see the discussion later), as well as further searches for suitable nuclei in the rare-earth region, might change the picture.

The variety of the possible physics programme and the diversity of the possible experimental scenarios are reflected in a variety of independent approaches to the development of both detector and accelerator technologies. The following sections provide an overview of the current efforts. As better knowledge becomes available both from the physics and the technology sides, we expect these efforts to focus on a more limited set of prioritized options.

**3.2 Neutrino detector studies**

The key to future discoveries in neutrino physics is likely to reside in the ability to produce extremely massive detectors while retaining enough information to eliminate backgrounds and provide precision measurements. In addition, it is necessary to study the indispensable near detectors and instrumentation of the primary beam line, be it the β-beam or neutrino factory storage ring, or the superbeam decay tunnel.

### 3.1.1 A megaton water Cherenkov detector

This is the baseline detector concept for a low-energy neutrino experiment with the combination of a β-beam and a superbeam. The designs currently under study follow generations of successful detectors. The performance of Super-Kamiokande has been widely studied and simulated, providing a basis for an extrapolation in detector mass by an order of magnitude. The studies are being carried out in the context of a dedicated international effort focused by the NNN workshop series. Three detector designs are being carried out, namely Hyper-Kamiokande in Japan, UNO in the US, and MEMPHYS in Europe. The critical issues are: i) Is an underground cavern hosting a megaton of water feasible? ii) How an affordable photo-sensor coverage can be provided? iii) What is the cost of such a project? A preliminary answer has been provided for the UNO project and more recently for MEMPHYS, where a recent civil engineering pre-study has allowed a first cost estimate [22].

Much study is still needed, and the key to an affordable detector will be R&D on large area, high quantum-efficiency photodetectors, which is currently being carried out in those three regions of the world. Studies are also needed of practical engineering solutions for the inner detector and its lining, and of protection against natural radioactivity.

### 3.1.2 Segmented magnetic detectors

These form the main line of study for the Neutrino Factory. The most studied Neutrino Factory detector so far has been the *magnetized iron detector*. The MINOS detector (5.6 kiloton) and the MONOLITH and INO projects (30 kiloton) are examples of the concept, for which the following questions can naturally be asked: Up to which mass can one extrapolate? Can one increase the sampling so as to lower the interesting muon detection threshold, i.e. the momentum for which a muon can be separated from hadron backgrounds, and its charge reliably determined?

Much of the progress has come from the application of the low-Z tracking calorimetry technique recently developed for NOVA. The design of magnetized iron toroids with radii up to 10 m can be extrapolated from MINOS. On the other hand, the NOVA liquid-scintillator technology, or the MINERVA solid-scintillator technology, either with APD or SiPM readout (as in the case of the T2K ND280 fine-grain detector), would allow the readout of fine-grain sensitive detectors of the corresponding size. An iron-scintillator detector of 100 kiloton with magnetized iron plates of 1 cm thickness seems feasible, and the simulation of such a detector has begun. A well-defined R&D project would be the construction and beam test of a small module of such a detector with a variety of readout schemes; in parallel, an engineering study of the full-size detector would be required.

In order to reach the full physics capabilities of a neutrino factory, detectors capable of detecting all the channels described in Table 3 are necessary, requiring in particular the capability of detecting taus and electrons and identifying their charges, for which *fully active magnetized detectors* are being studied.

*Table 3: Detection channels available at a neutrino factory*

| $\mu^+ \rightarrow e^+ \nu_e \bar{\nu}_\mu$ | $\mu^- \rightarrow e^- \nu_\mu \bar{\nu}_e$ | Reaction | |
|---|---|---|---|
| $\bar{\nu}_\mu \rightarrow \bar{\nu}_\mu$ | $\nu_\mu \rightarrow \nu_\mu$ | CC | Disappearance |
| $\bar{\nu}_\mu \rightarrow \bar{\nu}_e$ | $\nu_\mu \rightarrow \nu_e$ | CC | Appearance: 'platinum' channel |
| $\bar{\nu}_\mu \rightarrow \bar{\nu}_\tau$ | $\nu_\mu \rightarrow \nu_\tau$ | CC | Appearance (atmospheric oscillations) |
| $\nu_e \rightarrow \nu_e$ | $\bar{\nu}_e \rightarrow \bar{\nu}_e$ | CC | Disappearance |
| $\nu_e \rightarrow \nu_\mu$ | $\bar{\nu}_e \rightarrow \bar{\nu}_\mu$ | CC | Appearance: 'golden' channel |
| $\nu_e \rightarrow \nu_\tau$ | $\bar{\nu}_e \rightarrow \bar{\nu}_\tau$ | CC | Appearance: 'silver' channel |
| $\nu \rightarrow \nu_s$ | $\bar{\nu} \rightarrow \nu_s$ | NC | Global disappearance, for sterile-neutrino search |

One possibility based on the NOVA development would be to surround a *fully-active scintillator detector* with coils providing a moderate magnetic field of around 0.4 T. The goal is here the good detection and sign determination of low-energy electrons with energies up to a few GeV, so as to study the 'platinum channel' $\nu_\mu \rightarrow \nu_e$. An optimization of the detector density and a design and cost study of such a large magnet would be required.

The use of an *Emulsion Cloud Chamber (ECC)* has been proposed to detect the tau channel. The model is the ECC of OPERA, which consists of a multiple sandwich of lead and nuclear emulsion sheets. For operation with a neutrino factory, the emulsion target area would need to be surrounded by a large magnet, and the lead–emulsion sandwich would have to be upgraded with a series of gaps with light material for precise momentum measurement, and with scintillator to provide the time stamp of the events. This would be particularly important if the neutrino factory is run with both signs of muons simultaneously, separated by a gap of about 100 ns. The first simulations of such a detector have revealed that not only would the Magnetized ECC (MECC) perform very efficiently for tau leptons, but also for the charge identification of muons and electrons. The physics reach of such a detector seems impressive, but it is necessary to quantify the maximum mass affordable in terms of scanning power and costs. In this case also, a very large air-core magnet would be needed.

Finally, a *very large liquid-argon TPC* with a mass ranging from 10 to 100 kiloton would be a powerful instrument, owing to its excellent event reconstruction capabilities. Versions of the concept exist for super-beams, β-beams and a neutrino factory. In the latter case, the need for a magnetic field constitutes an additional challenge, but the combined possibilities of precise reconstruction of the event kinematics and electron charge identification in the few GeV range constitute strong motivations. The typical magnet needed has a stored energy of the same order as that of the CMS solenoid. A large liquid-argon detector would also allow a programme of experiments in astroparticle physics and proton decay to be conducted.

This concept is being studied by active collaborations in Europe and the USA. Recent highlights of the liquid-argon R&D programme include the construction and operation of a 300-ton prototype in Pavia and the operation of a 10-l prototype in a magnetic field. Preparations are under way to test the feasibility of very long drifts using a 5 m vertical tube, followed by readout in the gaseous phase, using, for example, large gas electron-multiplier (LEM) devices. The issue of delivering the necessary high voltage (up to 2 MV in the case of a 20 m drift length) is also being addressed, and a first look has been given to the possibility of using high-$T_c$ superconductors for the coils. At the same time, engineering studies have begun, in collaboration with industry specialized in the construction of very large tanks of liquid methane.

This work should continue with a more elaborate and detailed industrial design of the large underground (deep or shallow) tank and the details of the detector instrumentation. Finally, the study of logistics, infrastructure and safety issues related to underground sites should also progress, in view of the two typical possible geographical configurations: a tunnel-access underground laboratory and a vertical-access (mine-type) underground laboratory.

It is felt that a necessary step, both from the point of view of the development of the technique and because it would provide valuable physics results, will be the construction of a liquid-argon near detector for one of the ongoing neutrino beams. Such a proposal exists for the T2K experiment, and plans have been made for a NUMI off-axis detector. The typical masses of these projects amount to about 100 tons, i.e. between one per cent and one per mille of the ultimate goal.

In order to demonstrate the potential of liquid-argon detectors in magnetic field, a beam test of a moderate size prototype (1×1×3 m$^3$) inside a magnetic field seems the natural R&D path.

### 3.1.3 Near detector and beam instrumentation
While the design and construction of the large far detectors will be a considerable technological challenge, the near detectors and beam instrumentation will have to provide basic information for the extraction of the physics results, such as neutrino cross sections, neutrino event properties and determination of the neutrino beam flux and composition. Present experiments aim at disappearance measurements or at the appearance of a new signal over background, which is sufficient to fold neutrino cross sections and flux in an extrapolation from the near detector to the far one. Future experiments, on the contrary, will aim at a precise determination of appearance probabilities and will require precise knowledge of the flux of the initial-flavour neutrinos and of the final-flavour cross sections separately. This is a considerable challenge and, for the first time, understanding of neutrino fluxes with per mille precision is being discussed. In addition, there is considerable neutrino physics, valuable in its own right, that can be performed using the near detectors at a neutrino factory, thanks to the extraordinarily large flux.

**3.1.4 Absolute flux monitoring for the neutrino factory and β-beam**
Two methods have been identified to do this: absolute knowledge of the flux from beam instrumentation, and normalization to known processes in the near detector.

The absolute flux can be derived from the known decays and the numbers of decaying particles. The requirements for a safe determination of the flux are then: i) precise counting of the number of particles circulating in the ring; ii) precise determination of the beam polarization (for the neutrino factory); iii) precise determination of the beam direction, angular divergence, energy and energy spread. These require specific instrumentation of the storage ring and place requirements on the beam optics. Beam monitoring with beam-current transformers, ring-imaging Cherenkov and a polarimeter has been discussed in the ECFA/CERN Studies of a European Neutrino Factory Complex. The actual implementation needs to be studied and prototypes of the concepts will need to be defined.

To validate the good knowledge of the flux, a necessary condition is the availability of a reference process analogous to the Bhabha process in $e^+e^-$ annihilation. Thanks to the impressive intensity of the flux in the near detectors, neutrino interactions on electrons ($\nu e^-$ to $\nu e^-$ or $\nu \mu^-$) offer this possibility, provided a suitable detector could be built to perform this measurement with the required accuracy. The conceptual design of such a detector needs to be developed and a simulation performed: prototyping may also be necessary, and work on this subject has started.

It is clear that such possibilities are absent for conventional hadron decay beams, and that the use of the full statistical power of the superbeam will hinge on hadroproduction experiments, with a precision that remains to be demonstrated.

**3.1.5 Cross-section measurements and the physics of neutrino interactions**
The requirements of neutrino-oscillation experiments capable of measuring a CP asymmetry will include a precise knowledge of exclusive processes and cross sections, for the appearance channel and for the backgrounds. It will be the task of near detectors to establish those.

For the low-energy superbeam and β-beam option, the emphasis will be on the understanding of nuclear target and muon mass effects that, for instance, affect the $\nu_e/\nu_\mu$ cross-section ratio differently for neutrinos and antineutrinos. At slightly higher energies, pion production is also sensitive to these effects, and precise measurements as functions of the ν energy may be necessary. A survey of the theoretical and experimental knowledge is being undertaken within the framework of the NUINT conferences [23].

For the neutrino factory, the simultaneous availability of all flavours in the same detectors offers many advantages. The study of purely leptonic processes and of neutral-current production is of interest for precision tests of the Standard Model. The backgrounds to the various oscillation channels will need to be studied carefully, in particular charm production. There are many choices for a detector technology that could be implemented. A liquid-argon TPC in a magnetic field would be able to carry out most of the near-detector programme. More conventional scintillator technology (similar to MINERVA), a scintillating fibre tracker or a gas TPC (as in the T2K near detector) would also be able to perform cross-section and flux control measurements. However, it seems likely that only silicon or emulsion detectors can achieve the necessary spatial resolution to perform the charm measurements needed to determine the background for the oscillation search. These options should be studied further, and a programme of R&D should be established.

## 3.3 Detector R&D strategy

It is clear from the above discussion that a significant amount of R&D on detector technology will be necessary to reduce the costs of such large detectors and to establish their technical feasibility.

On a worldwide basis, dedicated R&D for future neutrino detectors is a small effort, essentially stemming from ongoing neutrino programmes. Some of the technologies emphasized above are presently being tested in the T2K experiment, namely the water Cherenkov, a near detector (ND280) based on a magnetized fine-grain detector, and there are studies for a possible liquid-argon TPC in the future 2 km near detector. The ECC is clearly a development of the OPERA detector in the CNGS programme, and the liquid-argon TPC is a development from ICARUS. The dedicated R&D necessary for the development of future detector projects is being evaluated at the moment in the framework of the ISS.

CERN currently contributes the magnet to T2K, and the work of HARP could be continued by measurements of particle production from the T2K target, which would clearly be an investment for the future. Several European groups are actively considering participation in NOνA. CERN has considerable expertise in the technology and construction of very large detectors and magnets, and could have a considerable impact. The availability at CERN of a test beam area with a magnet of ~ 0.5 T over a volume of the typical size mentioned above (1×1×3 m$^3$) would constitute common good for the neutrino community for the testing of magnetized detector prototypes.

## 3.4 Accelerator R&D

It is well known that extensive R&D is required for either the superbeam/β-beam combination or the neutrino factory.

Both the superbeam and the neutrino factory would require a 4 MW proton driver such as the SPL accompanied by an accumulator ring. An additional proton bunch-compressor ring is necessary for the neutrino factory. A rapid cycling synchrotron with energy in the range 5–15 GeV would also be a possibility.

The target area is very demanding. The physics of the target itself is site-independent and is currently being addressed via the MERIT experiment at nTOF, in which a liquid-mercury target is exposed to high-intensity proton pulses in a high magnetic field. However, issues related to safety, remote handling, and containment must imperatively be studied locally and a design produced of the target area.

In the case of the neutrino factory, muon beam cooling provides a large factor in the flux. R&D on cooling is currently under way in the MUCOOL programme in the US, and a full-scale test of a cooling section is the ultimate goal of the MICE experiment under way at RAL. New, promising, ideas for cooling both in energy and transverse dimensions (emittance exchange) are at present under study. They involve a helicoidal magnetic channel and high-pressure $H_2$-filled RF cavities. After successful prototyping, and if matching optical solutions are found, they could be tested experimentally in MICE.

The acceleration of muons involves either recirculating linacs or non-scaling FFAGs to accelerate muons rapidly to about 20 GeV. Either of these options would require further R&D projects. An important aspect would be the availability of suitable RF sources and RF cavities at the frequencies required by the muon bunch sizes (200–800 MHz). The design of the storage ring requires taking into account the physics constraints and the needs to monitor the beam intensity and divergence.

Neutrino factory R&D is currently being assessed via the International Scoping Study. This is expected to lead, in 2007, to a proposal to the EU to fund a design study that should deliver a Conceptual Design Report in 2010-12.

In parallel, the R&D for a $\gamma = 100$ β-beam is being addressed within the EURISOL framework, with conclusions to be available in 2010. The substantial investment in a combined superbeam and β-beam facility, including the megaton detector, will be defined better in the same time frame. The β-beam would require less than half a megawatt of proton beam power, but would in addition require a new low-energy storage and acceleration stage in the CERN accelerator complex. The prospects of achieving the required ion intensities are currently under study and require R&D.

A new method for production of nuclei in inverse kinematics has recently been proposed [24]. The nuclei are produced in a compact low-energy ring, which acts as both an ionization-cooling ring for the circulating stable ion beam and as a source for unstable nuclei through nuclear reactions in a gas target. The nuclei are then extracted with foils using an ISOL-like set-up. This method looks very promising and opens new possibilities for β-beams. New isotopes will become available, widening the scope of the β-beam concept.

The present baseline isotope, $^{18}$Ne, can probably be produced in sufficient quantities to realize the physics goal of the γ = 100 β-beam, which is at present a major problem for the ongoing study of a γ = 100 β-beam facility. We strongly recommend further evaluation and prototyping for this concept.

We note that all the possible major future options for neutrino physics at CERN could be based on the SPL or an equivalent proton accelerator intermediate between the LINAC4 and the new PS. The γ = 100 β-beam would benefit from an upgraded PS that would address space-charge limitations. In addition, a higher energy β-beam would be facilitated by the higher energy of the SPS+. We also note that the replacement currently proposed for the PS is not a high-power machine in the MW range, and that no such machine is currently envisaged by the PAF Working Group [2,3].

# 4 Other physics

## 4.1 Rare kaon decays

*Prospects at CERN*
Experiments on kaon physics have been a traditional strength of CERN, and have obtained many important results on CP violation, in particular. The chain of CERN proton accelerators provides the possibility of pursuing a very competitive and cost effective program at present, and future upgrades of the proton complex would allow the community to plan for next-generation experiments.

The most interesting subject addressed by current kaon experiments is the study of rare kaon decays. In the future, the highest priorities in kaon decay experiments are studies of the K → πνν decay modes, both neutral and charged [25]. These flavour-changing neutral decays are particularly interesting because they are precisely calculable loop processes in the Standard Model, which may well get significant corrections from extensions such as supersymmetry, and complement measurements of B mesons, as seen in the unitarity triangle shown in Fig. 12. The experimental constraints on the unitarity triangle will improve during the next few years, thanks to improved B factory measurements, data from the Fermilab Tevatron, and the initial data from LHCb, but the overall picture is not expected to change qualitatively on this time scale.

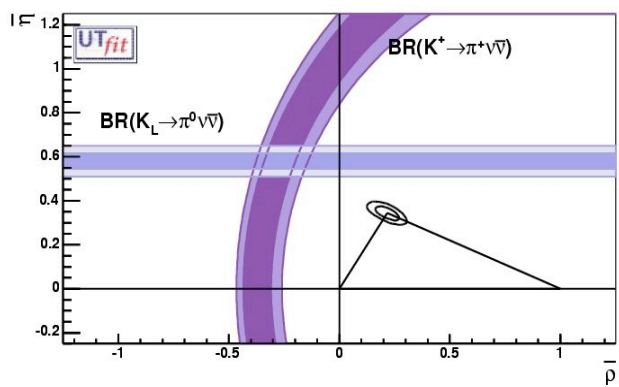

*Figure 12 – The potential impacts of K→πνν measurements compared to a present fit to the Cabibbo-Kobayashi-Maskawa unitarity triangle, which are largely derived from measurements of B decays [6].*

Projects to measure K → πνν decays in the US have been cancelled, but the neutral mode may be measured at KEK and JPARC by the E391a experiment. The P-326 (NA48/3) proposal aims at measuring the charged mode at the SPS. The latter would start taking data in 2009/2010, with the objective of obtaining around 80 events by about 2012, assuming the Standard Model branching ratio. This would provide a constraint on flavour mixing independent of those being provided by B mesons. The NA48/3 apparatus could also be modified to obtain about 30 $K_L \to \pi^0 \nu\nu$ events if the branching ratio is similar to that predicted in the Standard Model, and a substantial fraction of the SPS proton intensity is used to produce neutral kaons. In addition, if the apparatus' tracker is retained, it could also be used to measure the $K_L \to \pi^0 e^+ e^-$ and $K_L \to \pi^0 \mu^+ \mu^-$ modes. These measurements could be completed by about 2016.

Follow-up measurements with greater accuracy would be very important if any of these first-generation experiments found a possible discrepancy with the Standard Model prediction based on B physics measurements. This would require giga-Hz kaon rates, for instance a sequel to P-326 using a separated kaon beam originating from a 4 MW 50 GeV proton beam. This could obtain over 1000 $K^+ \to \pi^+ \nu\nu$ events per year, assuming the Standard Model branching ratio.

It is important to stress that the crucial parameter for the quality of a kaon experiment is the machine duty cycle, which should be as close to 100% as possible. The duration of the flat top for slow extraction may be limited by the power consumption of the warm magnets. The energy of the proton beam is not the primary concern for the design of a rare kaon decay experiment, provided it is high enough to produce kaons efficiently.

*The flavour-physics context*
The central goal of any new initiative in flavour physics should be to uncover physics beyond the Standard Model (BSM), and to probe the flavour structure of any BSM physics that may soon be discovered at the high-energy frontier provided by experiments at the LHC. The guiding principle for judging the potential of new flavour experiments should be to compare their expected physics output with the broader knowledge expected to be available when the new initiatives will be ready to collect data. For instance, in assessing the current P-326 proposal, one should take into account the knowledge expected by the time P-326 will be ready, namely around 2010. By that date the potential of the current generation of B factories will be fully exploited and the LHC experiments should have provided the first round of precise measurements in B physics. In general, observables that are both precisely calculable in the Standard Model (in other words that are not affected by hadronic uncertainties) and are sensitive to New Physics will remain interesting at that stage: examples from K and B decays via flavour-violating neutral interactions are given in the table below.

*Table 4: Prospective theoretical and experimental errors in flavour physics*

| Observable | Theoretical error | Likely error or sensitivity (experiment) |
|---|---|---|
| $B(K^+ \to \pi^+ \nu\nu)$ | ~6% | 75% (BNL) <br> 15% (2 years of P-326) |
| $B(K^0_L \to \pi^0 \nu\nu)$ | ~3% | ~$10^{-9}$ (E391a) |
| $B(B \to X_s \gamma)$ | ~10% (5% @NNLO?) | ~5% (Belle+Babar) |
| $B(B \to X_s l^+ l^-)$ | ~13% | ~20% (LHCb) |
| $A_{FB}(B \to K^{(*)} l^+ l^-)$ | ~15% | ~30% (LHCb) |
| $B(B \to (K^{(*)} \rho, \omega)\gamma)$ | ~25% | ~40% (LHCb) |
| $B(B_s \to \mu^+ \mu^-)$ | ~30% | SM (ATLAS/CMS) |
| $B(B \to K^* l^+ l^-)$ | ~35% | ~13% (LHCb) |

We see that the interpretations of the prospective measurements of the branching ratios for $K^+ \to \pi^+ \nu\nu$ and $K^0_L \to \pi^0 \nu\nu$ are unlikely to be clouded by the theoretical errors, which is not necessarily the case for $B \to X_s\gamma$ and $B \to K^* l^+ l^-$. In a couple of years, the current generation of B factories should deliver measurements of the $B \to X_s\gamma$ branching ratio at the 5% level but, to exploit such precision, the SM prediction needs to improve via a complete NNLO calculation of the QCD corrections, which is currently under way. Improvements of the theoretical calculation of $B \to K^* l^+ l^-$ decay will also be necessary, if the likely LHCb experimental precision is to be matched. Measurements of $\sin 2\beta$, $B_s$ mixing, $B \to \tau\nu$ decay and $B_s \to \mu^+\mu^-$ decay are now approaching the level where they put pressure on the Standard Model and begin to constrain possible extensions such as superymmetry. Significant improvements can be expected with coming data from the B factories, the Fermilab Tevatron and LHCb. These data and any hints they might provide for new physics should be taken into account in considering any extension of P-326 beyond its initial measurement of $K^+ \to \pi^+ \nu\nu$. However, the physics interest of this first phase is already clear.

*Experimental considerations*
Experiments reaching the Standard Model sensitivity for the charged decay mode $K^+ \to \pi^+ \nu\nu$ have already been performed, and three candidate events were published by the BNL experiments E787/E949. To be useful, the next generation of experiments will have to make a precise (i.e. 10% or better) measurement of the branching ratio in order to constrain the theoretical prediction significantly. Whilst the first-generation experiments were performed using kaon decays at rest, it is likely that in order to improve the experimental acceptance and the overall environmental cleanliness, the next studies will be performed with kaon decays in flight. A great advantage of this decay mode with respect to the neutral one is that the initial state can be completely determined. A disadvantage is that the two-pion decay of the charged kaon is not suppressed by CP symmetry as in the case of the neutral kaon, and therefore particular care has to be taken to suppress possible backgrounds originating when the two-body kinematics is badly reconstructed and photons from the $\pi^0$ decay are lost.

In-flight proposals can be divided into separated and unseparated experiments, depending on the purity of the incoming kaon beam. Separated beams have a small contamination of other particles such as pions and protons. The advantage is that that the burden on the beam tracking detectors is given mostly by useful kaons. The disadvantage is that to separate kaons from other particles efficiently, the momentum of the secondary beam cannot be larger than 20 GeV. On the other hand, unseparated secondary beams put a burden on the beam tracker but allow a better background rejection. Given the recent developments in tracking detectors, high-rate unseparated beams have become a cost-effective alternative to studying this decay. This is the road chosen in the P-326 proposal for the SPS.

The situation for the corresponding neutral decay mode $K^0_L \to \pi^0 \nu \nu$ is rather different, since present experiments are still far from being sensitive to the Standard Model prediction, and the measurement of this theoretically clean but elusive decay mode is an experimental challenge. Since the neutrino–antineutrino pair cannot be measured directly, it would be good to know as much as possible about the incoming neutral kaon: this could in principle be determined very accurately in the process $e^+e^-$ to $\phi$ to $K_S K_L$, but the only viable method to produce a large enough amount of $K^0_L$ is using a proton beam hitting a fixed target. The energy of the proton beam is not critical, as long as it is far enough above threshold to make kaons with good efficiency. It is possible to determine the momentum of the neutral kaon by a time-of-flight (TOF) measurement, but only if its momentum is less than about 1 GeV. This was the method proposed by the KOPIO experiment, which unfortunately will not be built. The TOF technique is very expensive in the use of protons because, in order to make low-energy kaons, the neutral beam needs to be extracted at a large production angle. This implies that, in order to collect a sufficient flux of neutral kaons, a very wide beam acceptance needs to be considered. In turn, the useful acceptance of practical detectors tends to be very small.

Currently, the only method being pursued to study $K^0_L \to \pi^0 \nu \nu$ is the so-called pencil-beam technique. Pioneered by KTeV at Fermilab, it consists in making neutral kaons together with many many neutrons and photons, and taking the neutral beam at a small production angle. The acceptance of the beam is then kept small so that the beam direction can be kept as a kinematical constraint in the form of a minimum transverse momentum cut. This is useful, for instance, to reject $\Lambda \to n\pi^0$ decays, where the neutron escapes detection and the $\pi^0$ mimics the final state under study.

If the kaon momentum is unknown, the only handle available is to measure fully the $\pi^0$ in the final state and veto any other particle using a hermetic detector. This method was proposed by KAMI at Fermilab, but was not considered viable by a technical review committee. Nevertheless, the same method is being applied in a step-by-step approach by the KEK E391a experiment at the 12 GeV PS, which is also proposed for further studies at the J-PARC hadron facility.

## 4.2 Fixed-target physics with a muon beam

COMPASS has an interesting programme of polarized muon scattering, in particular, and the COMPASS detector is well suited for a continuing physics programme with CERN's unique muon beam after 2010. The COMPASS collaboration is proposing an extension of its present approved programme, aiming at the measurement of generalized parton distribution functions via deeply–virtual Compton scattering and hard meson-exchange processes. The modified apparatus could be prepared between 2007 and 2009 and take data between 2010 and 2015 [26].

## 4.3 Fixed-target heavy-ion physics [27]

In addition to relativistic heavy-ion collisions at the LHC (discussed earlier), there may be an interesting physics opportunity for a continuation of CERN's fixed-target programme using heavy-ion beams. A critical point in the quark–hadron phase diagram is thought to be accessible to fixed-target heavy-ion experiments at the SPS, although its signatures are uncertain. It is expected that data would be needed over a range of energies, with good luminosity. The possible step-wise suppression of $J/\psi$ production would be one objective, as well as charm production studies. In order to establish a baseline for nuclear absorption, studies of both proton–ion and light-ion collisions would be desirable.

This programme could be addressed using an NA60-like dimuon spectrometer. Other research topics would include studies of low-mass lepton pairs (in the $\rho$ region, to understand better the possible effects of chiral-symmetry restoration), the observed enhancement of intermediate-mass dimuon production (which is thought to be thermal, rather than due to charm production), and the search for fluctuations in particle yields in narrow ranges of effective temperature and baryon chemical potential (which could be a signal for the critical point of the QCD phase transition). A possible scenario would include runs in the period from 2010 to 2013, including Pb–Pb, Cu–Cu and Pb–Be collisions. If the SPS+ were to become available, runs with Pb–Pb or U–U at the highest possible beam energy would also be interesting. A continuation of the NA60 programme to study further aspects of charm and charmonium production in proton collisions is also under consideration.

We note that the FAIR facility at GSI also plans a fixed-target programme to explore the QCD phase transition. When this is better defined, it will be necessary to assess the value that could be added by a continuation of the SPS fixed-target programme.

## 4.4 Nuclear physics

The evolution of experimental nuclear physics towards ever more complex experiments, combined with the need for large beam power for the generation of highly exotic isotopes, will in the future require a laboratory infrastructure for nuclear physics similar to that required for high-energy physics. Nuclear physics is closer to applied sciences than particle physics, but the subject is motivated by a drive to explore the unknown, and is in its nature a fundamental science. This, together with the potential synergies at the infrastructure level, generates large common ground between the subjects.

CERN currently has a forefront research programme in nuclear physics, based on ISOLDE and nTOF. The on-line isotope separation technique perfected at ISOLDE provides a uniquely broad range of isotopes with fast changeovers, which in turn provide many interesting physics opportunities. As seen in Fig. 13, these include studies of *nuclear structure*, e.g. of light nuclei with extended neutron haloes, and of the breakdown of the shell model far from stability, with a transition from order to quantum chaos in nuclear spectroscopy; studies in *nuclear astrophysics*, e.g. of the *s*- and *r*-processes as well as light nuclei; *probes of the Standard Model and its possible extensions*, e.g. measurements of β transitions that bear upon Kobayashi–Maskawa unitarity; and *applications* to solid-state physics and the life sciences. nTOF also provides unique opportunities, e.g. for measuring neutron cross sections and for fission studies. A community of about 600 nuclear physicists is currently using these facilities at CERN.

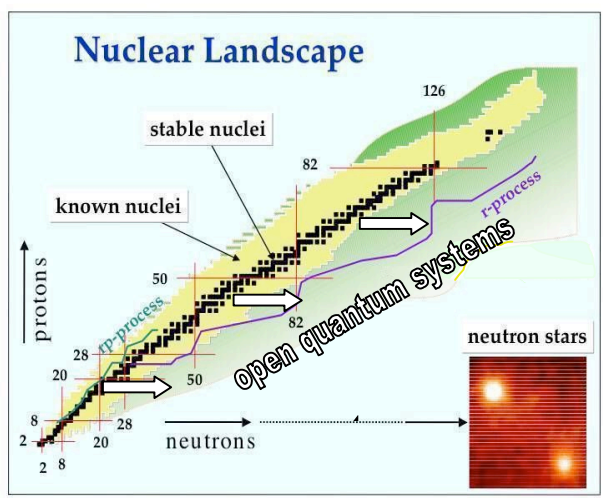

*Figure 13 – Outline of the nuclear landscape, highlighting open research areas such as studies of processes relevant to astrophysics, and the breakdown of the shell model.*

The ISOLDE community is integrated within the EURISOL initiative [28], which is planning two phases for developing new facilities. In the first phase, it is proposed to upgrade the present REX-ISOLDE facility to accelerate radioactive isotopes up to 10 MeV/u, the HIE-ISOLDE project [29], which would enable a large range of open nuclear-structure questions to be addressed. The present ISOLDE beam energy of 3 MeV/u limits Coulomb excitation experiments to the lower half of the nuclear chart. An increase to 5.5 MeV/u would enable Coulomb excitation of all the nuclei produced at ISOLDE, making it possible to study their low-energy structure and shape evolution. The energy increase to 10 MeV/u provided by HIE-ISOLDE would enable transfer reactions to be studied also, giving much more detailed structure information, e.g. on the evolution of shells. The option of deceleration to very low energies will permit the study of cross sections for astrophysics in the energy range of special interest for the *s*- and *r*-processes.

The HIE-ISOLDE extension of the present ISOLDE programme could be completed by 2012, and would benefit from Linac4 and an upgrade of the PS booster to increase its frequency. (We note in passing that nTOF would also benefit from Linac4.) The proposed upgrade programme at ISOLDE, based on stepwise improvements of the existing facility during the annual shutdown periods, promises an interesting cutting-edge nuclear physics programme, if it is realized within the proposed schedule. The competition from other proposed next-generation facilities in Europe will start in earnest only towards the latter part of the next decade, and a sustained aggressive target and ion-source development programme at ISOLDE should continue to assure a leading position for ISOLDE even during this era.

In the second phase, EURISOL proposes [28] a major new low-energy proton accelerator operating at 1 GeV, providing by 2015 100-kW beams on three separate targets, one of which could yield the $^6$He and $^{18}$Ne needed for a β-beam facility. There would also be a 5 MW beam incident on a spallation neutron target. The facility would deliver radioactive ion beams of up to 150 MeV/u from all target stations from a common linac.

The linac will provide high-intensity radioactive ion beams orders of magnitude more intense than those currently available, while having an emittance comparable to stable-beam accelerators and having a continuously tunable energy. Its technical specifications, in terms of RIB intensities, energy range, beam purity and quality will be unique worldwide. For a large range of radionuclides (in particular neutron-rich nuclei), the primary production yields of EURISOL are superior to those from facilities such as FAIR, that use the in-flight method.

EURISOL will provide a facility for research that addresses the major challenge of the fundamental understanding of nuclear structure in terms of the underlying many-body interactions between hadrons. Descriptions of nuclei having more than a few nucleons are semi-phenomenological in origin and cannot be applied reliably to nuclei far from stability. It is therefore crucial to measure the properties of nuclei at the extremes of stability, such as the evolution of shell structure, T = 0 and T = 1 pairing, and collective phenomena such as halo effects and pygmy resonances. EURISOL also aims at understanding the Universe through its history of stellar activity and galaxy formation, where nuclear reactions play essential roles. In particular, in the violent maelstrom of explosive processes such as novae, x-ray bursters and supernovae, the heavy elements are made in complex networks of reactions (*r*- and *rp*-processes) on unstable nuclei and via β decays. To understand these processes quantitatively and identify the astronomical sites where they occur requires a wealth of information on unstable nuclei. EURISOL, with its broad range of beams, will permit the study of many of the key reactions. Further applications of EURISOL to fundamental tests of the Standard Model, to the application of unique probes for surface science and condensed matter studies, and to other fields can be found in the report for EURISOL prepared within the RTD contract in the 5th framework [30]. EURISOL will have applications that benefit society in many different areas and have strong impact on other fields of science, as illustrated in the NuPECC report [31].

ISOLDE and its possible extensions make possible an impressive range of nuclear physics, and also potential synergies with the rest of the CERN programme, e.g., with neutrino physics. The concept of the β-beam has been made possible by ISOLDE, and the proton driver for EURISOL might also serve as a driver for a neutrino superbeam or a neutrino factory. In this case, a global optimization of its energy should be undertaken.

The EURISOL community is estimated at 1000 physicists, a large fraction of the nuclear physicists thought to be needed in Europe in the future to support the many applications of nuclear physics on which our society depends, e.g. radio-pharmacology, materials science, and energy production. The construction of EURISOL would require new resources that would go beyond the scale of investments traditionally made in nuclear physics at CERN.

The recent proposal to produce nuclei in inverse kinematics in a compact low-energy ring [24] may have wider applications beyond the β-beams discussed earlier, which should also be studied further.

## 4.5 Antiproton physics

The tradition at CERN in the use of low-energy antiprotons originates from the SPS proton-antiproton collider times, when it was realized that the abundant production of antiprotons offered the unique possibility to decelerate and cool them to provide beams of unprecedented intensity and purity. The Intersecting Storage Rings (ISR) was used for colliding-beam and internal-target experiments with antiprotons from the Antiproton Accumulator (AA). Later a small facility, LEAR, for experiments with cooled antiprotons, was added to the CERN complex. The tradition continues today with the Antiproton Decelerator (AD). The antiproton programme at CERN has been varied and included topics such as charmonium formation, nucleon–antiproton annihilations, the study of CP violation in neutral kaon decays, the spectroscopy of antiprotonic atoms, and the formation of antihydrogen. This work is currently being continued by the ATRAP, ALPHA and ASACUSA collaborations.

The main scientific interest in future antiproton experiments at CERN lies in studies of antihydrogen and tests of CPT invariance. CPT violation is forbidden in the usual theoretical framework of quantum field theory, so its discovery would be a major event of considerable fundamental impact. To reach the very high precision needed for relevant CPT tests, trapping antihydrogen in a magnetic field is needed. The coming years will be dominated by technical developments by the ATRAP, ALPHA and ASACUSA collaborations, who are taking different approaches to the challenging tasks of trapping and cooling antihydrogen. Once this R&D programme has been successfully completed, hopefully by about 2009, it will become possible to envisage a future round of experiments directed towards precise tests of the CPT theorem and measurements of the gravitational acceleration of antimatter.

Another possible topic for future antiproton studies at CERN is a programme of measurements of cross sections for the elastic and inelastic scattering of low-energy (0 to 100 MeV) antiprotons on protons and light nuclei such as deuterium, tritium and helium [32]. These are of interest for cosmology, as they may have played important roles in modifying primordial light-element abundances in models with unstable heavy particles, and could be studied with the ASACUSA setup. Also interesting would be the corresponding antineutron interactions, whose feasibility using antineutrons produced via charge exchange from antiprotons remains to be studied.

We note that CERN's pioneering work with antiprotons and antihydrogen has not only generated scientific discoveries, but also garnered for CERN considerable public attention, sometimes unexpected but most of it welcome. This aspect should also be considered when assessing the Laboratory's future antiproton programme. The coming years will also see the advent of the FAIR project, which will provide competition for CERN in antiproton physics.

## 4.6 Electric dipole moments

CERN should be open to other suggestions for exploiting the rich proton accelerator infrastructure present and envisaged here. The ongoing Workshop on Flavour Physics in the LHC Era has brought up the intriguing suggestion to make a highly competitive measurement of the CP-violating deuteron electric dipole moment with an accuracy $\sim 10^{-29}$ e.cm using a 1.5 GeV/c storage ring [33]. The observation of a deuteron electric dipole moment would be a major discovery, and such a project would complement CERN's continuing experiments on CP violation in K and B mesons, as well (possibly) as neutrinos.

This idea could be implemented at CERN by sending polarized deuterons into the LEIR ring for accumulation and bunching, before transferring them into a dedicated ring for experimental runs that would each last several hours. *A priori*, there is no obvious incompatibility with using LEIR to provide heavy ions for the LHC, so this interesting project might have good synergies with CERN's approved programmes, and would not require extensive effort from the accelerator departments. However, there are other possible locations for this project, such as BNL, and its competitivity with the neutron electric dipole experiments at ILL and PSI should be assessed carefully.

## 4.7 Muon physics

The anomalous magnetic moment of the muon and its electric dipole moment are already prominent topics in muon physics. In the future, the headline question in muon physics may become the possible existence of flavour-violating µ → e transitions [34]. This would be the case, in particular, if the LHC reveals new physics at the TeV scale, such as supersymmetry. In conjunction with the lepton-flavour violation seen in neutrino physics, this would suggest that charged-lepton flavour violation might also be detectable. The discovery of flavour-violating µ → e transitions would be a suitable high-level goal for CERN. We also note that intense muon beams incident on a fixed target may provide an opportunity to look for μ → τ transitions in deep-inelastic scattering.

An experiment with sensitivity to μ → eγ decay at the level of $10^{-13}$ is being prepared at PSI, whereas a proposed search for anomalous μ → e conversion on nuclei at BNL has recently been turned down. If μ → eγ decay is found at PSI, follow-up searches for μ → eee and μ → e anomalous conversion will become enhanced priorities, but the interest of the latter would be strong even if the PSI experiment does not detect μ → eγ decay. A programme in muon physics is planned for JPARC, but not yet funded. We note that R&D and prototyping for a possible muon physics facility is currently being conducted by the PRIME collaboration on a site-independent basis. This project should be ready by 2010 for a decision to be made on whether to site it at JPARC or elsewhere.

The construction of a high-intensity, low-energy proton driver for either neutrino and/or nuclear physics would provide CERN with a scientific opportunity to host a world-leading muon-physics facility.

## 4.8 Electron–hadron collider

There have been two expressions of interest in colliding electrons at an energy around 70 GeV with a proton and/or heavy-ion beam in the LHC (LHeC). The QCD Explorer suggestion is to use an electron beam accelerated by a CLIC module, which would be an interesting synergy with a possible CERN Lepton Facility [35]. More recently, the idea of reinstalling an electron ring in the LHC tunnel has been resurrected, with the prospect of attaining a significantly higher luminosity of $10^{33}$ cm$^{-2}$s$^{-1}$ [36]. We note that this project could use ILC superconducting RF cavities, but would also require the construction of a new lepton injector chain, since that used for LEP has now been dismantled and partly reused for other activities, such as CTF3. We are conscious that both these LHeC options would require very considerable resources, and their physics potentials will need to be evaluated carefully.

They would open up a completely new kinematic range for deep-inelastic scattering, as seen in Fig. 14, which would certainly have great interest for QCD studies. The LHeC would also have capabilities for detecting new physics, such as leptoquarks or supersymmetry. Fig. 15 gives one example, namely the cross section for associated production of selectron–squark pairs at the LHeC. This might be particularly interesting in a scenario where low-mass sparticles are discovered in the initial phase of LHC running. The better-constrained kinematics of electron–proton scattering would enable improved measurements of the sparticle properties to be made.

The LHeC project could also provide electron–ion collisions, which would have interest for relativistic heavy-ion physics. It would serve to probe parton saturation and the possible existence of the colour-glass condensate with more clarity than would have been possible with the ion–ion or proton–ion collisions at the LHC. Initial results from these previous programmes would provide key physics input for such an electron–ion project, which would collide electrons with energies around 70 GeV with LHC ion beams at around 5.5 TeV/u. These energies are far beyond the capabilities of the eRHIC project at BNL, which envisages colliding 5 to 10 GeV electrons with 250 GeV protons and 100 GeV/u nuclei, in the period up to 2020.

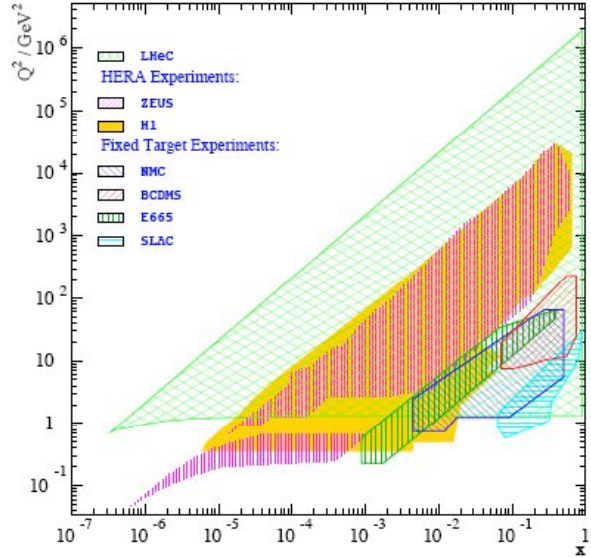

*Figure 14: The kinematic range open to the LHeC, compared with that accessible to previous facilities for lepton–proton collisions [36].*

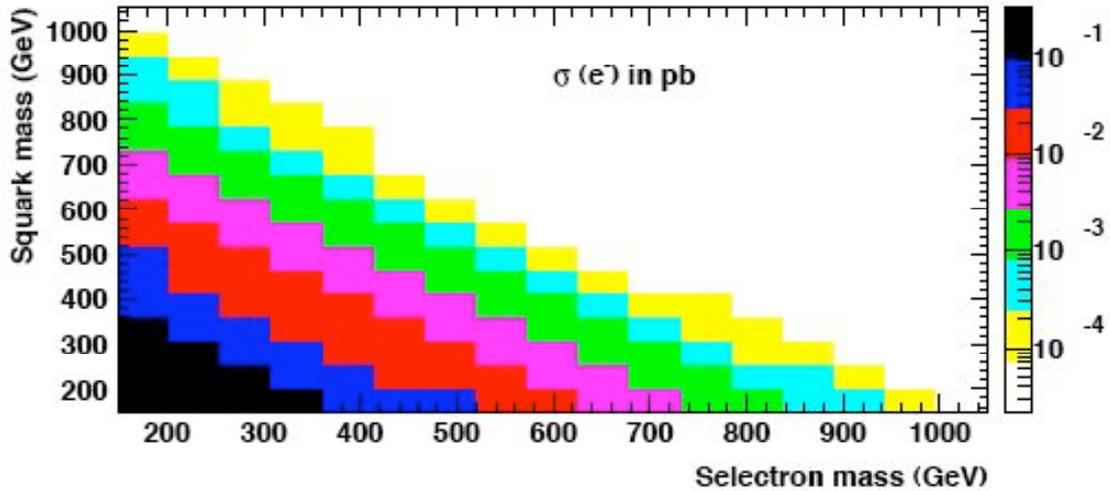

*Figure 15: An example of the possible cross section for associated production of slectrons and squarks at the LHeC in a representative supersymmetric scenario [36].*

# 5 Conclusions

In this report we have expanded the discussion of the physics drivers for possible future proton accelerators at CERN that was given in our initial report [1]. In the coming years, the ordered priorities we advocate are full exploitation of the LHC, together with preparation for possible luminosity and energy upgrades [2]; preparation for a possible next-generation neutrino facility at CERN [3]; and optimizing the use of CERN infrastructures via unique fixed-target and low-energy programmes in areas such as kaon physics and nuclear physics.

We have outlined the corresponding detector R&D priorities, including the four LHC experiments, neutrino detectors and kaon physics.

It will be necessary to keep under review the physics drivers for CERN's future proton accelerator options. In parallel, it will be necessary to compare the physics opportunities offered by proton accelerators with those available at a linear $e^+e^-$ collider in the ILC or CLIC energy range.

This continuing review will draw on early experience with LHC running and the initial physics results, as well as changes in the physics perspectives due to experimental advances elsewhere, theoretical developments, and technological breakthroughs.